\documentclass[aps,pre,twocolumn,superscriptaddress,showpacs]{revtex4}
\usepackage{amssymb}
\usepackage{epsfig}
\usepackage{amsmath}
\usepackage{times}
\begin{document}

\title{Transportation dynamics on networks of mobile agents}

\author{Han-Xin Yang}\email{hxyang@mail.ustc.edu.cn}
\affiliation{Department of Modern Physics, University of Science and
Technology of China, Hefei 230026, China}

\author{Wen-Xu Wang}
\affiliation{School of Electrical, Computer and Energy
Engineering, Arizona State University, Tempe, Arizona 85287, USA}

\author{Yan-Bo Xie}
\affiliation{Department of Modern Physics, University of Science
and Technology of China, Hefei 230026, China}

\author{Ying-Cheng Lai}
\affiliation{School of Electrical, Computer and Energy
Engineering, Arizona State University, Tempe, Arizona 85287, USA}
\affiliation{Department of Physics, Arizona State University,
Tempe, Arizona 85287, USA}

\author{Bing-Hong Wang}\email{bhwang@ustc.edu.cn}
\affiliation{Department of Modern Physics, University of Science and
Technology of China, Hefei 230026, China} \affiliation{ Research
Center for Complex System Science, University of Shanghai for
Science and Technology and Shanghai Academy of System Science,
Shanghai, 200093 China}

\begin{abstract}
Most existing works on transportation dynamics focus on networks of
a fixed structure, but networks whose nodes are mobile have become
widespread, such as cell-phone networks. We introduce a model to
explore the basic physics of transportation on mobile networks. Of
particular interest are the dependence of the throughput on the
speed of agent movement and communication range. Our computations
reveal a hierarchical dependence for the former while, for the
latter, we find an algebraic power law between the throughput and
the communication range with an exponent determined by the speed. We
develop a physical theory based on the Fokker-Planck equation to
explain these phenomena. Our findings provide insights into complex
transportation dynamics arising commonly in natural and engineering
systems.

\end{abstract}

\date{\today}


\maketitle

\section{Introduction} \label{sec:intro}

Transportation processes are common in complex natural and
engineering systems, examples of which include transmission of data
packets on the Internet, public transportation systems, migration of
carbon in biosystems, and virus propagation in social and
ecosystems. In the past decade, transportation dynamics have been
studied extensively in the framework of complex networks
\cite{ASG:2001,KYHJ:2002,TB:2004,TTR:2004,MB:2004,
PLZY:2005,SG:2005,CLLF:2005,DYMB:2006,LRMHSA:2010}, where a
phenomenon of main interest is the transition from free flow to
traffic congestion. For example, it is of both basic and practical
interest to understand the effect of network structure and routing
protocols on the emergence of congestion
\cite{ZLPY:2005,WWYXZ:2006,ZLRH:2007,GKL:2008,yang2008,TLLH:2009,
LHJWCW:2009,TM:2009,XWLHH:2010}. Despite these works, relatively
little attention has been paid to the role of {\em individual
mobility}. The purpose of this paper is to address how this mobility
affects the emergence of congestion in transportation dynamics.

The issue of individual mobility has become increasingly
fundamental due to the widespread use of ad-hoc wireless
communication networks. The issue is also important in other
contexts such as the emergence of cooperation among individuals
\cite{HY:2009} and species coexistence in cyclic competing games
\cite{RMF:2007}. Recently, some empirical data of human movements
have been collected and analyzed \cite{HBG:2006,GHB:2008}. From
the standpoint of complex networks, when individuals (nodes,
agents) are mobile, the edges in the network are no longer fixed,
requiring different strategies to investigate the dynamics on such
networks than those for networks with fixed topology. In this
paper, we shall introduce an intuitive but physically reasonable
model to deal with transportation dynamics on such
mobile/non-stationary networks. In particular, we assume in our
model that communication between two agents is possible only when
their geographical distance is less than a pre-defined value, such
as the case in wireless communication. Information packets are
transmitted from their sources to destinations through this
scheme. To be concrete, we assume the physical region of interest
is a square in the plane, and we focus on how the communication
radius and moving speed may affect the transportation dynamics in
terms of the emergence of congestion. Our main results are the
following. Firstly, we find that congestion can occur for small
communication range, limited forwarding capability and low mobile
velocity of agents. Secondly, the transportation throughput
exhibits a hierarchical structure with respect to the moving speed
and there is in fact an algebraic power law between the throughput
and the communication radius, where the power exponent tends to
assume a smaller value for higher moving speed. To explain these
phenomena in a quantitative manner, we develop a physical theory
based on solutions to the Fokker-Planck equation under initial and
boundary conditions that are specifically suited with the
transportation dynamics on mobile-agent networks. Besides
providing insights into issues in complex dynamical systems such
as contact process, random-walk theory, and self-organized
dynamics, our results will have direct applications in systems of
tremendous importance such as ad-hoc communication networks
\cite{MAHN4,MAHN6,MAHN7}.

In Sec. \ref{sec:model}, we describe our model of mobile agents in
terms of the transportation rule and the network structure. In Sec.
\ref{sec:numerics}, we present numerical results on the order
parameter, the critical transition point and the average hopping
time. In Sec. \ref{sec:theory}, a physical theory is presented to
explain the numerical results. A brief conclusion is presented in
Sec. \ref{sec:conclusion}.

\section{Transportation rule and network structure of mobile agents}
\label{sec:model}

\begin{figure}
\begin{center}
\epsfig{figure=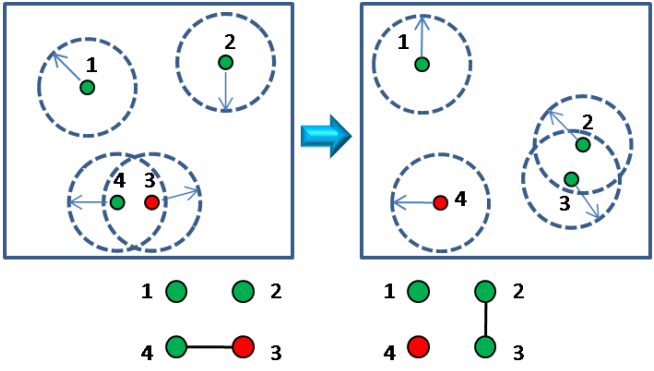,width=\linewidth} \caption{(Color online)
Schematic illustration of transportation process among mobile
agents. The dashed circles denote the communication range, the
arrows denote the moving directions, and each agent is specified by
a number. Agents without data packets are in green (light gray) and
agents holding data packets are in red (dark gray). In the left
panel, agent 3 and 4 can communicate with each other, where the
former holds a packet. Agent 3 delivers the packet to agent 4 and
then all agents move, as shown in the right panel. After the
movements, all agents randomly set new directions. There can be more
than one agent in the communication range of any agent. The relevant
communication networks are shown in the bottom. }
\label{fig:schemeic}
\end{center}
\end{figure}

In our model, $N$ agents move on a square-shaped cell of size $L$
with periodic boundary conditions. Agents change their directions of
motion $\theta$ as time evolves, but the moving speed $v$ is the
same for all agents. Initially, agents are randomly distributed on
the cell. After each time step, the position and moving direction of
an arbitrary agent $i$ are updated according to
\begin{equation}
x_{i}(t+1)=x_{i}(t)+v\cos\theta_{i}(t),
\end{equation}
\begin{equation}
y_{i}(t+1)=y_{i}(t)+v\sin\theta_{i}(t),
\end{equation}
\begin{equation}
\theta_{i}(t)=\Psi_{i},
\end{equation}
where $x_{i}(t)$ and $y_{i}(t)$ are the coordinates of the agent at
time $t$, and $\Psi_{i}$ is an $N$-independent random variable
uniformly distributed in the interval $[-\pi,\pi]$. Each agent has
the same communication radius $a$. Two agents can communicate with
each other if the distance between them is less than $a$. At each
time step, there are $R$ packets generated in the system, with
randomly chosen source and destination agents, and each agent can
deliver at most $C$ packet (we set $C=1$ in this paper) toward its
destination. To transport a packet, an agent performs a local search
within a circle of radius $a$. If the packet's destination is found
within the searched area, it will be delivered directly to the
destination and the packet will be removed immediately. Otherwise,
the packet is forwarded to a randomly chosen agent in the searched
area. The queue length of each agent is assumed to be unlimited and
the first-in-first-out principle holds for the queue. The
transportation process is schematically illustrated in
Fig.~\ref{fig:schemeic}.

The communication network among the mobile agents can be extracted
as follows. Every agent is regarded as a node of the network and a
wireless link is established between two agents if their
geographical distance is less than the communication radius $a$.
Due to the movement of agents, the network's structure evolves
from time to time. The network evolution as a result of local
mobility of agents is analogous to a locally rewiring process. As
shown in Fig.~\ref{fig:schemeic}, nodes 1 and 2 are disconnected
while node 3 and node 4 are connected at time $t$. At time $t+1$,
nodes 1 and 2 depart from each other and become disconnected while
nodes 3 and 4 approach each other and establish a communication
link. Note that the mobile process does not hold the same
number of links at different time, which is different from the
standard rewiring process where the number of links is usually fixed.

\begin{figure}
\begin{center}
\epsfig{figure=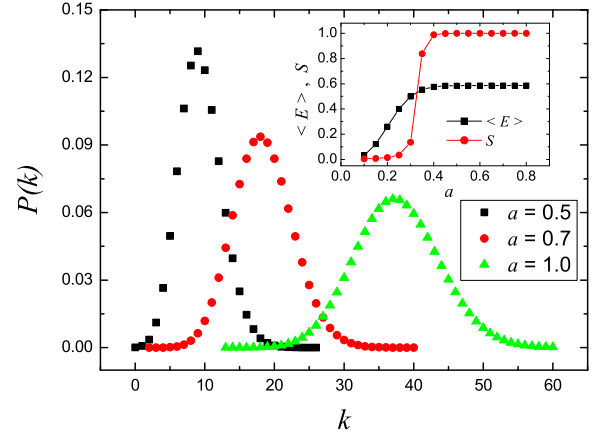,width=\linewidth} \caption{(Color online)
The proportion of nodes $P(k)$ as a function of degree $k$ for
different communication radius $a$. The inset shows that the
relative size $S$ of the largest connected component and the
clustering coefficient $\langle E \rangle$ as a function of
communication radius $a$. The number of agents is 1200 and the size
of the square region is $L=10$.} \label{fig:degree}
\end{center}
\end{figure}

We define an agent's degree at a specific time step as the number of
links at that moment. Figure~\ref{fig:degree} shows that the degree
distribution of networks of mobile agents exhibits the Poisson
distribution:
\begin{equation}
P(k)=\frac{e^{-\langle k \rangle}\langle k \rangle^{k}}{k!},
\end{equation}
where $k$ is the degree, $P(k)$ is the proportion of nodes with
degree $k$ and $\langle k \rangle$ is the average degree of network.
As shown in Fig.~\ref{fig:degree}, the average degree $\langle k
\rangle$ increases as the communication radius $a$ increases and the
peak value of $P(k)$ decreases as $a$ increases. We also investigate
the relative size of the largest connected component and the
clustering properties of the network in terms of the clustering
coefficient. The relative size $S$ of the largest connected
component is defined as
\begin{equation}
S=\frac{S_{1}}{N},
\end{equation}
where $S_{1}$ and $N$ is the size of the largest connected component
and the total network respectively. The clustering coefficient
$E_{i}$ for node $i$ is defined as the ratio between the number of
edges $e_{i}$ among the $k_{i}$ neighbors of node $i$ and its
maximum possible value, $k_{i}(k_{i}-1)/2$, i.e.,
\begin{equation}
E_{i}=\frac{2e_{i}}{k_{i}(k_{i}-1)}.
\end{equation}
The average clustering coefficient $\langle E \rangle$ is the
average of $E_{i}$ over all nodes in the network. The insert of
Fig.~\ref{fig:degree} shows that $S$ and $\langle E \rangle$
increase as the communication radius $a$ increases. In particular,
when the value of $a$ exceeds a certain value. e.g., $0.4$, high
values of $S$ and $\langle E \rangle$ is attained. We also note that
the motion speed does not influence the statistical properties of
the communication network. In general, the communication network
caused by limited searching area and mobile behavior is of
geographically local connections associated with Poisson
distribution of node degrees and dense clustering structures.

\section{Numerical results} \label{sec:numerics}

\begin{figure}
\begin{center}
\epsfig{figure=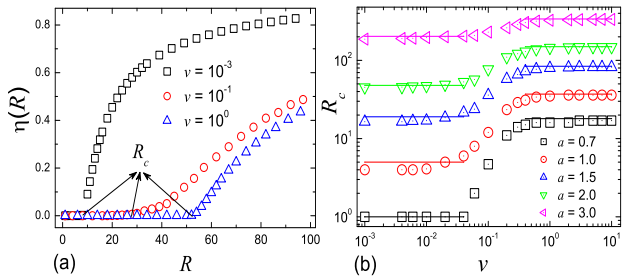,width=1.06\linewidth}\caption{(Color
online) (a) Order parameter $\eta(R)$ versus $R$ for different
values of speed $v$ and (b) dependence of the critical $R_c$ on $v$
for different communication radius $a$. The number of agents is 1200
and the size of the square region is $L=10$. The delivery capacity
$C$ of each agent is set to be unity. In (a), $a=1.2$ and $\eta(R)$
is obtained by averaging over $10^{5}$ time steps after disregarding
$2\times 10^{4}$ initial steps as transients. The results in both
(a) and (b) are obtained by an ensemble average of 20 independent
realizations. The lines in (b) are theoretical prediction from
Eqs.~(\ref{7}) and (\ref{10}). } \label{fig:order}
\end{center}
\end{figure}

To characterize the throughput of a network, we exploit the order
parameter $\eta$ introduced in Ref. \cite{ASG:2001}:
\begin{equation}
\eta(R)=\lim_{t\rightarrow\infty} \frac{C}{R}\frac{\langle\Delta
N_{p}\rangle}{\Delta t},
\end{equation}
where $\Delta N_{p}=N_{p}(t+\Delta t)-N_{p}(t)$, $\langle \cdot
\cdot \cdot \rangle$ indicates the average over a time window of
width $\Delta t$, and $N_{p}(t)$ represents the total number of data
packets in the whole network at time $t$. As the packet-generation
rate $R$ is increased through a critical value of $R_{c}$, a
transition occurs from free flow to congestion. For $R \leq R_{c}$,
due to the absence of congestion, there is a balance between the
number of generated and that of removed packets so that $\langle
\Delta N_{p}\rangle= 0$, leading to $\eta(R)= 0$. In contrast, for
$R>R_{c}$, congestion occurs and packets will accumulate at some
agents, resulting in a positive value of $\eta(R)$. The traffic
throughput of the system can thus be characterized by the critical
value $R_c$ which is on average the largest number of generated
packets that can be handled at each time without congestion.

\begin{figure}
\begin{center}
\epsfig{figure=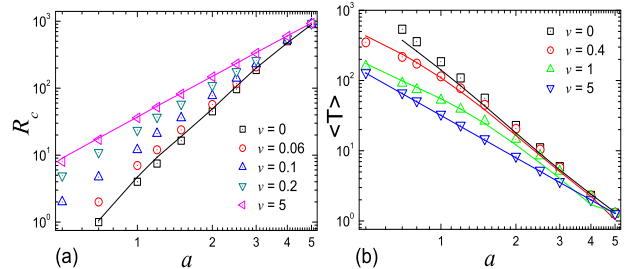,width=1.06\linewidth} \caption{(Color
online) (a) Critical value $R_c$ as a function of the communication
radius $a$ for different values of $v$. The lines are theoretical
predictions from Eqs.~(\ref{7}) and (\ref{10}). (b) Average hopping
time $\langle T \rangle$ as a function of $a$ for different values
of $v$. The theoretical curves are obtained from
Eqs.~(\ref{4.9})-(\ref{eq:v3}). For $v=0.4$ and $v=1$, $f=1$ is
used. The average time is computed in the free flow state. Each data
point is obtained by averaging over 20 independent realizations.}
\label{fig:RcT}
\end{center}
\end{figure}

Figure~\ref{fig:order}(a) exemplifies the transition in the order
parameter $\eta(R)$ from free flow to congestion state at some
critical value $R_c$. We find that $R_c$ depends on both the
moving speed $v$ and the communication radius $a$.
Figure~\ref{fig:order}(b) shows the dependence of $R_c$ on $v$ for
different values of $a$. We observe a hierarchical structure in
the dependence. Specifically, when $v$ is less or larger than some
values, $R_c$ remains unchanged at a low and a high value,
respectively, regardless of the values of $v$. The transition
between these two values of $R_c$ is continuous. The hierarchical
structure can in fact be predicted theoretically in a quantitative
manner (to be described). Figure \ref{fig:RcT}(a) shows the
dependence of $R_c$ on $a$ for different values of $v$, which
indicates an algebraic power law: $R_c \sim a^\beta$, where
$\beta$ is the power-law exponent. We find that the power law
holds for a wide range of $a$ and the exponent $\beta$ is
inversely correlated with $v$. For example, for $v = 0$, $\beta
\approx 3.5$ but for large values of $v$, say $v = 5$, we have
$\beta \approx 2$. When $a$ reaches the size of the square cell,
$R_c$ is close to $N$ as every agent always stays in the searching
range of all others and almost all packets can arrive at their
destinations in a single time step.

To gain additional insights into the dependence of $R_c$ on the
parameters $a$ and $v$ so as to facilitate the development of a
physical theory, we explore an alternative quantity, the average
hopping time $\langle T \rangle$ in the free flow state which, for
a data packet, is defined as the number of hops from its source to
destination. As we will see, $\langle T \rangle$ can not only be
calculated numerically, it is also amenable to theoretical
analysis, providing key insights into the theory for $R_c$.
Representative numerical results for $\langle T \rangle$ are shown
in Fig.~\ref{fig:RcT}(b). We see that for large $v$, $\langle
T\rangle$ scales with $a$ as $a^{-2}$ and, as both $v$ and $a$ are
increased, $\langle T\rangle$ decreases.

\section{Theory} \label{sec:theory}

We now present a physical theory to explain the power law
behaviors associated with $\langle T \rangle$ and then $R_c$. A
starting point is to examine the limiting case of $v=0$, where
$\langle T \rangle$ can be estimated analytically. In particular,
assume that a particle walks randomly on an infinite plane. There
are many holes of radius $a$ on the plane. Holes form a phalanx
and the distance between two nearby holes is $L$. The particle
will stop walking when it falls in a hole. The underlying
Fokker-Planck equation is
\begin{equation}
\frac{\partial P(\textbf{r},t)}{\partial
t}=[A\nabla^{2}+U(\textbf{r})]P(\textbf{r},t),
\end{equation}
where $P(\textbf{r},t)$ is the probability density function of a
particle at location ${\textbf r}$ at time $t$, $A$ is the diffusion
coefficient, $U(\textbf{r})$ is the potential energy,
$U(\textbf{r})=-\infty$ inside holes and $U(\textbf{r})=0$ outside
holes, and $P(\textbf{r},t)=0$ inside holes and $P(\textbf{r},t)>0$
outside holes. Making use of solutions to the eigenvalue problem:
\begin{equation} \label{4.2}
[A\nabla^{2}+U(\textbf{r})]\Phi_{n}(\textbf{r})=-\lambda_{n}\Phi_{n}(\textbf{r}),
\end{equation}
where $\Phi_{n}(\textbf{r})$ is the normalized eigenfunction and
$\lambda_{n}$ is the corresponding eigenvalue, we can expand
$P(\textbf{r},t)$ as
\begin{equation}
P(\textbf{r},t)=\sum_{n=1}^{\infty}c_{n}e^{-\lambda_{n}t}\Phi_{n}(\textbf{r}),
\end{equation} where $c_{n}=\int
P(\textbf{r},0)\Phi_{n}(\textbf{r})d\textbf{r}$ and the initial
probability density $P(\textbf{r},0)$ is distributed over a region
of a typical size $a$. The probability that a particle still walks
at time $t$ is:
\begin{equation}
Q(t)=\int
P(\textbf{r},t)d\textbf{r}=\sum_{n=1}^{\infty}c_{n}d_{n}e^{-\lambda_{n}t},
\end{equation}
where $d_{n}=\int\Phi_{n}(\textbf{r})d\textbf{r}$. Since the $n=1$
term is dominant, we have
\begin{equation}
Q(t)\approx c_{1}d_{1}e^{-\lambda_{1}t},
\end{equation}
which gives the average hopping time as
\begin{equation}
\langle T \rangle \approx \frac{1}{\lambda_{1}}.
\end{equation}
Since
\begin{equation}
\Phi_{1}(x,y)=\Phi_{1}(x+L,y)=\Phi_{1}(x,y+L),
\end{equation}
the infinite-plane problem can be transformed into a problem on
torus:
\begin{eqnarray} \label{4.5}
\nabla^{2}\Phi_{1}(r)=-h^{2}\Phi_{1}(r), \ \ \mbox{for} \ \
a<r<b,\nonumber \\ \Phi_{1}(r=a)=0, \Phi_{1}'(r=b)=0,
\end{eqnarray}
where $b=L/\sqrt{\pi}$ and
\begin{equation}
\Phi_{1}(r)=B_{1}J_{0}(hr)+B_{2}N_{0}(hr).
\end{equation}
$J_{0}$ and $N_{0}$ are the first-kind and the second-kind Bessel
Function, respectively, and $B_{2}=-B_{1}J_{0}(ha)/N_{0}(ha)$. The
quantity $h$ can be obtained by
\begin{equation}
\frac{J_{0}'(hb)}{J_{0}(ha)}=\frac{N_{0}'(hb)}{N_{0}(ha)}.
\end{equation} Using
\begin{equation}
J_0(x) \approx 1-\frac{x^{2}}{4}+\frac{x^{4}}{64},
\end{equation}
\begin{equation}
N_0(x) \approx
\frac{2}{\pi}[(\ln{\frac{x}{2}}+0.5772)(1-\frac{x^{2}}{4}+\frac{x^{4}}{64})
+ \frac{x^{2}}{4} - \frac{3x^{4}}{128}],
\end{equation}
and combining Eqs. (\ref{4.2}) and (\ref{4.5}), we get
$\lambda_{1}=Ah^{2}$. For $\Delta t=1$, $\langle (\Delta r)^{2}
\rangle=a^{2}/2$, we have $A=\langle (\Delta r)^{2}
\rangle/4=a^{2}/8$ and $ \lambda_{1}=Ah^{2}=a^2h^2/8$. Finally, we
obtain $\langle T \rangle$ as
\begin{equation}
\langle T
\rangle\approx\frac{1}{\lambda_{1}}=\frac{8}{a^{2}h^{2}}.
 \label{4.9}
\end{equation}
For $v>0$, $P(\textbf{r},t)$ becomes zero in the area where a hole
moves and $Q(t)$ decays with time under two mechanisms: diffusion
at rate $\lambda_{1}$ and motion of holes at the rate
$\lambda_{w}$. Thus, we have
\begin{equation}
\langle T
\rangle\approx\frac{1}{f\lambda_{1}+\lambda_{w}},\label{eq:v2}
\end{equation}
where $f$ is the weighting factor ($0\leq f\leq1$) that decreases as
$v$ increases. Specifically, $f=1$ for small $v$ and $f=0$ for large
$v$. The quantity $\lambda_{w}$ is given by
\begin{eqnarray}
\hbox{(i)}\ \ \lambda_{w}&=&\frac{\int_{a}^{v_{s}+a}r\Phi_{1}(r)\arccos(\frac{r^{2}+v_{s}^{2}-a^{2}}{2rv_{s}})dr}{\pi\int_{a}^{b}r\Phi_{1}(r)dr},\nonumber \\
\hbox{(ii)}\ \ \lambda_{w}&=&\frac{\int_{v_{s}-a}^{v_{s}+a}r\Phi_{1}(r)\arccos(\frac{r^{2}+v_{s}^{2}-a^{2}}{2rv_{s}})dr}{\pi\int_{a}^{b}r\Phi_{1}(r)dr},\nonumber \\
\hbox{(iii)}\ \ \lambda_{w}&=&\frac{a^{2}}{b^{2}}=\frac{\pi a^{2}}{
L^{2}},\label{eq:lambda}
\end{eqnarray}
where $v_{s}=\sqrt{\langle
|\textbf{v}_{1}-\textbf{v}_{2}|^{2}\rangle}=\sqrt{2}v$, (i) is
valid for $0<v_{s}<2a<b-a$ or $0<v_{s}<b-a<2a$, (ii) is valid for
$2a<v_{s}<b-a$ and (iii) is for $v_{s}>b-a$. For large values of
$v$, agents are approximately well-mixed so that we can
intuitively expect the average time $\langle T \rangle$
to be determined by the inverse of the ratio of the agent's
searching area and the area of the cell:
\begin{equation}
\langle T \rangle \approx  \frac{L^{2}}{\pi a^{2}}. \label{eq:v3}
\end{equation}
The validity of this equation is supported by the fact that the
ratio of the two areas is equivalent to the ratio of the total
number of agents to the number of agents within the searching
area. The estimation of $\langle T \rangle$ for large $v$ is
consistent with the theoretical prediction from Eqs.~(\ref{eq:v2})
and (\ref{eq:lambda})(iii) by inserting $f=0$. The theoretical
prediction is in good agreement with simulation results, as shown
in Fig.~\ref{fig:RcT}(b).

With the aid of Eqs. (\ref{4.9}) and (\ref{eq:v3}) for $\langle T
\rangle$, we can derive a power law for $R_c$. In a free-flow
state, the number of disposed packets is the same as that of
generated packets $Rt$ in a time interval $t$. For $v=0$, the
number of packets $n_i$ passing through an agent is proportional
to its degree. This yields
\begin{equation}
n_i=\frac{R\langle T \rangle tk_i}{\Sigma_j k_j}=\frac{R\langle T
\rangle tk_i}{N\langle k \rangle},
\end{equation}
where $k_i$ is the degree of $i$, the sum runs over all agents in
the network, and $\langle k\rangle$ is the average degree of the
network. During $t$ steps, an agent can deliver at most $C_{i}t$
packets. To avoid congestion requires $n_{i}\leq C_{i}t$. If all
agents have the same delivering capacity $C$, the transportation
dynamics is dominated by the agent with the largest number of
neighbors and the transition point $R_{c}$ can be estimated by
\begin{equation}
\frac{R_{c}\langle T \rangle tk_{max}}{N\langle k\rangle}=Ct ,
\end{equation} where $k_{max}$ is the largest degree of the network.
Thus, for $v = 0$, we have
\begin{equation}
R_{c}=\frac{NC\langle k\rangle}{\langle T \rangle
k_{max}},\label{7}
\end{equation}
where $\langle k\rangle=N\pi a^2/L^2$. Since the degree distribution
follows the Poisson distribution: $P(k)=e^{-\langle k
\rangle}\langle k \rangle^{k}/k!$, the quantity $k_{max}$ can thus
be estimated by
\begin{equation}
\frac{e^{-\langle k \rangle}\langle k
\rangle^{k_{max}}}{k_{max}!}\simeq \frac{1}{N}.
\end{equation}
Inserting $\langle T\rangle$, $\langle k\rangle$
and $k_{max}$ into Eqs. (\ref{7}), we can calculate $R_c$ for low
moving speed $v$.

For large $v$, Eq. (\ref{7}) can also be applied but
$k_{max}=\langle k\rangle$ and $\langle T \rangle=L^{2}/\pi
a^{2}$. Hence, $R_{c}$ for large $v$ is given by
\begin{equation}
R_{c}=\frac{NC}{L^{2}}\pi a^{2}.\label{10}
\end{equation}
Equation (\ref{10}) indicates that $R_{c}$ scales with $a^{2}$,
which is in good agreement with simulation results shown in
Fig.~\ref{fig:RcT}(a).

\section{Conclusion} \label{sec:conclusion}

In conclusion, we have introduced a physical model to study the
transportation dynamics on networks of mobile agents, where
communication among agents is confined in a circular area of
radius $a$ and agents move with fix speed $v$ but in random
directions. In general, the critical packet-generating rate $R_c$
at which a transition in the transportation dynamics from free
flow to congestion occurs depends on both $a$ and $v$, and we have
provided a theory to explain the dependence. Our results yield
physical insights into critical technological systems such as
ad-hoc wireless communication networks. For example, the power
laws for the network throughput uncovered in this paper can guide
the design of better routing protocols for such communication
networks. From the standpoint of basic physics, our findings are
relevant to general dynamics in complex systems consisting of
mobile agents, in contrast to many existing works where no such
mobility is assumed.

\begin{acknowledgments}
This work is funded by the National Basic Research Program of China
(973 Program No.2006CB705500), the National Important Research
Project:(Study on emergency management for non-conventional happened
thunderbolts, Grant No. 91024026), the National Natural Science
Foundation of China (Grant Nos. 10975126 and 10635040), and the
Specialized Research Fund for the Doctoral Program of Higher
Education of China (Grant No. 20093402110032). WXW and YCL are
supported by AFOSR under Grant No. FA9550-10-1-0083.
\end{acknowledgments}


\begin{references}

\bibitem{ASG:2001} A. Arenas, A. D\'{\i}az-Guilera, and R. Guimer\`{a}, Phys. Rev. Lett.
\textbf{86}, 3196 (2001).

\bibitem{KYHJ:2002} B. J. Kim, C. N. Yoon, S. K. Han, and H. Jeong, Phys. Rev. E \textbf{65},
027103 (2002).

\bibitem{TB:2004} Z. Toroczkai and K. E. Bassler, Nature (London) {\bf 428}, 716
(2004).

\bibitem{TTR:2004} B. Tadi\'{c}, S. Thurner, and G. J. Rodgers, Phys. Rev. E \textbf{69},
036102 (2004).

\bibitem{MB:2004} M. A. de Menezes and A.-L. Barab\'asi, Phys. Rev. Lett. 92,
028701 (2004).

\bibitem{PLZY:2005} K. Park, Y.-C. Lai, L. Zhao, and N. Ye,
Phys. Rev. E 71, 065105 (2005)

\bibitem{SG:2005} B. K. Singh and N. Gupte, Phys. Rev. E \textbf{71},
055103(R) (2005).

\bibitem{CLLF:2005} V. Cholvi, V. Laderas, L. L\'{o}pez, and A.
Fern\'{a}ndez, Phys. Rev. E \textbf{71}, 035103(R) (2005).

\bibitem{DYMB:2006} B. Danila, Y. Yu, J. A. Marsh, and K. E. Bassler, Phys. Rev. E \textbf{74}, 046106 (2006).

\bibitem{LRMHSA:2010} G. Li, S. D. S. Reis, A. A. Moreira, S. Havlin, H. E. Stanley, and J. S. Andrade,
Jr., Phys. Rev. Lett. \textbf{104}, 018701 (2010).

\bibitem{ZLPY:2005} L. Zhao, Y.-C. Lai, K. Park, and N. Ye, Phys. Rev. E \textbf{71},
026125 (2005).

\bibitem{WWYXZ:2006}
W.-X. Wang, B.-H. Wang, C.-Y. Yin, Y.-B. Xie, and T. Zhou, Phys.
Rev. E \textbf{73}, 026111 (2006); W.-X. Wang, C.-Y. Yin, G. Yan,
and B.-H. Wang, {\it ibid.} \textbf{74}, 016101 (2006).

\bibitem{ZLRH:2007} H. Zhang, Z. Liu, M. Tang, and P. M. Hui, Phys. Lett. A \textbf{364}, 177
(2007).

\bibitem{GKL:2008} X. Gong, L. Kun, and C.-H. Lai, Europhys. Lett. \textbf{83}, 28001 (2008).
\bibitem{yang2008} H.-X. Yang, W.-X. Wang, Z.-X. Wu, and B.-H.
Wang, Physica A \textbf{387}, 6857 (2008).

\bibitem{TLLH:2009} M. Tang, Z. Liu, X. Liang, and P. M. Hui, Phys. Rev. E \textbf{80}, 026114
(2009).

\bibitem{LHJWCW:2009} X. Ling, M.-B. Hu, R. Jiang, R. Wang, X.-B. Cao, and Q.-S.
Wu, Phys. Rev. E \textbf{80}, 066110 (2009).

\bibitem{TM:2009} B. Tadi\'{c} and M. Mitrovi\'{c}, Eur. Phys. J.
B \textbf{71}, 631 (2009).

\bibitem{XWLHH:2010} Y.-H. Xue, J. Wang, L. Li, D. He, and B.
Hu, Phys. Rev. E \textbf{81}, 037101 (2010).


\bibitem{HY:2009} D. Helbing and W. Yu,
Proc. Natl Acad. Sci. USA {\bf 106}, 3680 (2009).

\bibitem{RMF:2007} T. Reichenbach, M. Mobilia, and E. Frey, Nature
(London) {\bf 448}, 1046 (2007).

\bibitem{HBG:2006} L. Hufnagel, D. Brockmann, and T. Geisel, Nature (London) {\bf 439}, 462 (2006).

\bibitem{GHB:2008} M. C. Gonz\'ales, C. A. Hidalgo, and A. L. Barab\'asi, Nature (London) {\bf 453}, 779
(2008).

\bibitem{MAHN4} L. Wang, C.-P. Zhu, and Z.-M. Gu, Phys. Rev. E \textbf{78}, 066107
(2008).

\bibitem{MAHN6} E. M. Royer and C.-K. Toh, IEEE Person. Commun.
\textbf{6}, 46 (1999).

\bibitem{MAHN7} T. Camp, J. Boleng, and V. Davies, Wirel. Commun. Mob. Comput.
\textbf{2}, 483 (2002).

\end{references}
\end{document}